\documentclass[iop,apjl,numberedappendix]{emulateapj}
\usepackage{amssymb}
\usepackage{amsmath}
\usepackage{graphicx}
\usepackage{natbib}
\usepackage{url}
\usepackage{paralist}
\usepackage{epsfig}

\def\cm{{\rm\thinspace cm}}

\def\erg{{\rm\thinspace erg}}

\def\K{{\rm\thinspace K}}

\def\s{{\rm\thinspace s}}

\def\yr{{\rm\thinspace yr}}

\def\pcmcu{\hbox{$\cm^{-3}\,$}}

\def\ergps{\hbox{$\erg\s^{-1}\,$}}

\def\pcmcu{\hbox{$\cm^{-3}\,$}}

\def\h18{\hbox{H1821$+$643\,}}

\def\bv{\hbox{Brunt-V\"ais\"al\"a\ }}
\def\vvel{\hbox{${\bf v}$}}

\slugcomment{ }

\shorttitle{Driving of ICM Turbulence by AGN}
\shortauthors{C.~S.~Reynolds et al.}

\begin{document}

\title{Inefficient Driving of Bulk Turbulence by Active Galactic Nuclei in a Hydrodynamic Model of the Intracluster Medium}

\author{Christopher~S.~Reynolds\altaffilmark{1,2}, Steven~A.~Balbus\altaffilmark{3}, and Alexander~A.~Schekochihin\altaffilmark{4,5} }

\altaffiltext{1}{Department of Astronomy, University of Maryland, College Park, MD 20742-2421, USA; chris@astro.umd.edu}
\altaffiltext{2}{Joint Space-Science Institute (JSI), College Park, MD 20742-2421, USA}
\altaffiltext{3}{Department of Physics, University of Oxford, Oxford OX1 3NP, UK}
\altaffiltext{4}{Rudolf Peierls Centre for Theoretical Physics, University of Oxford, Oxford OX1 3NP, UK}
\altaffiltext{5}{Merton College, Oxford OX1 4JD, UK}

\begin{abstract}
\noindent Central jetted active galactic nuclei (AGN) appear to heat the core regions of the intracluster medium (ICM) in cooling-core galaxy clusters and groups, thereby preventing a cooling catastrophe.  However, the physical mechanism(s) by which the directed flow of kinetic energy is thermalized throughout the ICM core remains unclear.  We examine one widely discussed mechanism whereby the AGN induces subsonic turbulence in the ambient medium, the dissipation of which provides the ICM heat source.  Through controlled inviscid 3-d hydrodynamic simulations, we verify that explosive AGN-like events can launch gravity waves ($g$-modes) into the ambient ICM which in turn decay to volume-filling turbulence.  In our model, however, this process is found to be inefficient, with less than 1\% of the energy injected by the AGN activity actually ending up in the turbulence of the ambient ICM.  This efficiency is an order of magnitude or more too small to explain the observations of AGN-feedback in galaxy clusters and groups with short central cooling times.   Atmospheres in which the $g$-modes are strongly trapped/confined have an even lower efficiency since, in these models, excitation of turbulence relies on the $g$-modes' ability to escape from the center of the cluster into the bulk ICM.  Our results suggest that, if AGN-induced turbulence is indeed the mechanism by which the AGN heats the ICM core, its driving may rely on physics beyond that captured in our ideal hydrodynamic model.  
\end{abstract}


%\keywords{accretion, accretion disks --- black hole physics --- galaxies: nuclei --- galaxies: Seyfert --- X-rays: binaries}

%%%%%%%%%%%%%%START PAPER%%%%%%%%%%%%%%%%%%%%%%%%%%%%%%%%%%%%%%%%%%%%

\section{Introduction}\label{intro}

It is now widely accepted that active galactic nuclei (AGN) regulate the growth of the most massive galaxies \citep[see review by ][ and references therein]{fabian:12a}.  Cooling-core clusters of galaxies provide a remarkably accessible view of these AGN feedback processes at work.  Most of the baryonic matter in galaxy clusters resides in an atmosphere of hot (temperatures $T\sim 10^7-10^8\K$) and tenuous (particle-number densities $n\sim 10^{-1}-10^{-3}\pcmcu$) intracluster medium (ICM).   Without any energy input, the observed atmospheres would radiatively cool on a timescale of $\sim 10^9\yr$ or less leading to a cooling catastrophe, levels of star formation far greater than those observed, and central galaxies that are much more massive than found in nature \citep{fabian:94a,mcnamara:07a}.    Instead, real clusters appear to be in a state of approximate thermal equilibrium, with some heat source appearing to balance the radiative cooling \citep{peterson:06a}.  Given the failings of models based on thermal conduction \citep{binney:81a,ruszkowski:02a,kim:03a,conroy:08a,bogdanovic:09a}, energy injection by an AGN hosted by the central galaxy appears to be the most viable source of this heat \citep{churazov:00a,reynolds:02a,churazov:02a,omma:04a}.

Indeed, the vast majority of cool-core clusters have central galaxies that host radio-loud AGN \citep{burns:90a}.  Furthermore, when sufficiently deep X-ray images are available, we invariably see evidence of interaction between the jets of these AGN and the ICM.  The most common signature of interaction are bi-polar ICM cavities that have been blown by jet-inflated cocoons \citep{fabian:00a,churazov:01a,heinz:02a,birzan:04a}.  In a hydrodynamic model, blowing well-defined low-density cavities before they are shredded by Rayleigh-Taylor instabilities requires supersonic initial expansion and hence we expect the presence of shocks within the ICM.   While strong shocks are never observed (implying that the strong shock phase may be short lived), there is evidence for weak shocks in a number of sources \cite[e.g., ][]{jones:02a,million:10a,randall:15a}, supporting the basic picture of supersonic jet-driven cavity expansion.

The basic energetics of the AGN heating of ICM cores also make sense.  In the Perseus cluster, for example, the X-ray luminosity emerging from within the cooling radius (usually defined as the radius within which the cooling time falls below 3\,Gyr) is $10^{45}\ergps$.  To create the inner, most well-defined pair of cavities requires in total approximately $4\times 10^{44}\ergps$, assuming adiabatic inflation.  However, they are bounded by weak shocks which add an additional $1.2\times 10^{45}\ergps$ to the energy budget \citep{graham:08a}.  Noting that these power estimates suffer from order-unity uncertainty owing to projection effects and uncertainties in the timescale of cavity inflation, we see that the AGN can clearly inject sufficient energy to heat the ICM core, provided that the process is reasonably efficient.  More generally, studies of samples of groups and clusters show that cavities are ubiquitous in systems with short central cooling times, and that between $1-20PV$ of energy per cavity (where $P$ is the pressure of the surrounding ICM and $V$ is the cavity volume) needs to be thermalized in order to offset the radiative cooling \citep{birzan:04a,panagoulia:14a}.  

However, going beyond these global energy arguments, the actual mechanism by which the AGN heating of the ICM occurs is not well understood.   Energy injection could be just that---direct injection of hot bubbles that mix with the cluster gas \citep{heinz:05a,hillel:15a}.  However, at least in the most powerful observed outbursts in intermediate temperature clusters, any such mixing must be slow (i.e., comparable to the cooling timescale) or else the cooling core would have been destroyed by the injected enthalpy \citep{pfrommer:12a,mcnamara:12a}.   Potentially of equal or greater importance is heating of the ICM by cosmic-ray (CR) protons as they diffuse/stream through the ICM \cite[e.g.,][]{loewenstein:91a,guo:08a,pfrommer:13a}.  The most likely source of CRs relevant for heating the cluster core is the jet-inflated cocoon/bubble itself and, hence, this is closely related to the bubble mixing issue already discussed.

Alternatively, the AGN energy injection can take the form of mechanical agitation that propagates into the ICM atmosphere initially in the form of waves, with the wave energy ultimately being thermalised \citep{churazov:04a}.   In this case, the heating is a secondary process.   There are multiple paths by which the thermalization might occur, depending upon the nature of the perturbation itself.  Sound waves propagate rapidly through the core and tend to steepen into weak shocks.  Buoyancy, or ``$g$-modes'' are much less compressive and propagate much less rapidly and extensively (indeed, as discussed in Section~\ref{theory}, they can be trapped within a finite radius).  These waves might be themalized directly if conduction or viscosity are sufficiently large \citep{ruszkowski:04a,fabian:05a,heinz:05a}, or indirectly on small scales via a turbulent cascade \citep{fujita:05a,schekochihin:06a,ensslin:06a,rebusco:06a,kunz:11a,zhuravleva:14a}.  The turbulent path could be particularly important when direct wave dissipation is inefficient.  None of these pathways to thermalization of mechanical energy has been systematically explored to see whether they are dynamically  viable.   

It is this important gap that our current work seeks to fill ---or begin to fill---by isolating and studying the process of turbulent dissipation of the g-modes generated by stirring in a cluster-like environment.  The turbulent heating picture recently received a major boost from a detailed analysis of X-ray surface brightness (SB) fluctuations in the Perseus and Virgo clusters as seen in deep {\it Chandra} images.   Interpreting the SB fluctuations as being due to turbulence\footnote{Of course, if there are any additional SB fluctuations arising from clumping/multiphase structure within the ICM, this analysis method would over-estimate the turbulent heating.}, \cite{zhuravleva:14a} determine the velocity power spectrum of the turbulence and proceed to estimate the volumetric turbulent dissipation as a function of radius.   They show that, within each radial bin that they probe, the derived turbulent dissipation approximately matches the observed radiative cooling.  The turbulence itself is envisaged to be due to the decay of $g$-modes driven into the bulk of the ICM through the action of AGN \citep[during the inflation and subsequent buoyant rise of the jet-blown cavities; ][]{omma:04a}.    Of course, the many other pathways by which wave thermalization might occur, including the role of magnetic fields, are a priori no less important, but lie outside the scope of the current work.

In this paper, we investigate the theoretical foundations of the AGN-driven turbulent heating picture.  We construct a toy-model atmosphere in which we then explore the driving of the hydrodynamic $g$-modes by explosive events (representing the AGN outbursts) and the subsequent decay of those modes into turbulence.  Our principal finding is that the AGN-like energy injection can indeed drive volume-filling turbulence throughout the ICM atmosphere but, within our particular model system, the transfer of energy from the AGN into the turbulent cascade is rather inefficient.  In our fiducial case, less than 1\% of the injected energy ends up as the kinetic energy of turbulence in the ambient ICM.  The majority of the injected energy is put into compressible modes (weak shocks and sound waves) that radiate out of the top boundary of our atmosphere, or are rapidly thermalized in the strongly turbulent cavity-debris zone. Furthermore, we find that atmospheres in which the $g$-modes are strongly trapped are even more inefficient because, in our models, driving turbulence in the bulk of the ICM is contingent on the $g$-modes being able to escape outwards from the center of the cluster. If the AGN-driven-turbulence picture gains strong support from future observational studies (for example, from direct X-ray spectroscopic measurements of large turbulent velocities by {\it Astro-H}), our work suggests that physics beyond ideal hydrodynamics is required to understand the driving of the turbulence. 

This paper is organized as follows.  In Section~\ref{theory}, we review some necessary theoretical concepts including the properties of $g$-modes in ICM atmospheres.  Our toy model is described in Section~\ref{simulations}.  Section~\ref{fiducial} describes the results for our fiducial parameter set with only weak $g$-mode trapping; the strong-trapping cases are treated in Section~\ref{trapping}.   The implications of our findings, as well as the limitations and shortcomings of our approach, are discussed in Section~\ref{discussion}.  Finally, Section~\ref{conclusions} draws together our conclusions.  

\section{Theoretical Preliminaries}\label{theory}

Before launching into a discussion of our model atmospheres and their dynamics, we first review in brief some relevant theoretical background.

%\subsection{$g$-modes}

We are concerned with the adiabatic hydrodynamic disturbances of an atmosphere about some hydrostatic equilibrium.  In linear theory, there are two families of modes corresponding to whether pressure or gravity acts as the restoring force.  The pressure modes are the familiar sound waves which propagate at the sound speed, $c_{\rm s}=\sqrt{\gamma P/\rho}$ where $P$ is the thermal pressure, $\rho$ the density, and $\gamma$ the ratio of specific heat capacities.    The pressure modes radiate out of any given region of interest on the sound crossing time.  

The gravity modes, or $g$-modes, are of central importance to our discussion.   To lowest order in the (small) Mach number of the velocity perturbations, they are incompressible, $\nabla\cdot \vvel=0$, where $\vvel$ is the velocity field.  This implies that the excitation of $g$-modes leads to the generation of vorticity, $\omega=\nabla\times\vvel$.  The evolution of the vorticity in the case of an ideal inviscid fluid subject to only conservative external forces is
\begin{equation}
\frac{D\omega}{Dt}=(\omega\cdot\nabla)\vvel-\omega(\nabla\cdot \vvel)+\frac{1}{\rho^2}\nabla\rho\times\nabla P,
\end{equation}
where $D/Dt=\partial/\partial t+(\vvel\cdot \nabla)$ is the convective derivative.  To linear order in the small velocities this gives
\begin{equation}
\frac{\partial\omega}{\partial t}=\frac{1}{\rho^2}\nabla\rho\times\nabla P.
\end{equation}
Hence, vorticity production is associated with departures between surfaces of constant density and those of constant pressure.  In the case of linear $g$-modes in an atmosphere, these departures are a consequence of the fact that $\rho={\rm constant}$ surfaces have first-order (in the Mach number of the velocity perturbations) corrugations relative to equipotentials whereas $P={\rm constant}$ surfaces only have second-order corrugations.  Given that the pressure gradient is then essentially in the vertical direction (as defined by the local gravitational field), the term $\nabla\rho\times\nabla P$ and hence the generated vorticity will lie principally in the horizontal plane.  We shall use this fact later in the paper to aid in the identification of $g$-modes. 

Gravity modes have the important property that they can be trapped.  To see this, we note that the dispersion relation of a local $g$-mode with a space-time dependence $\propto e^{i(\Omega t-{\bf k}\cdot {\bf r})}$ is
\begin{equation}
\Omega^2=\frac{k_h^2}{k^2}N^2.
\end{equation}
Here, $k=|{\bf k}|$, $k_h$ is the magnitude of the wavenumber projected onto the horizontal plane, $N$ is the \bv frequency given by
\begin{equation}
N^2=\frac{g(z)}{\gamma}\frac{\partial }{\partial z}\ln (P\rho^{-\gamma}),\\\label{eq:bv}
\end{equation}
and the gravitational field is defined to be ${\bf g}=-g(z)\hat{\bf z}$ ($\hat{\bf z}$ is the unit vector in the $z$-direction).  Since $k_h\le |{\bf k}|$, $g$-modes have a maximum possible frequency of $\Omega_{\rm max}=N$, at which point $k_h=|{\bf k}|, k_z=0$.  If the \bv frequency $N(z)$ is a decreasing function of height $z$, $g$-modes of a given frequency $\Omega$ will be trapped/confined below the height at which $N(z)=\Omega$. We shall see that $g$-mode trapping may have a significant effect on the ability of localized AGN driving to generate turbulence in the ambient medium.

\section{The Model and Computational Setup}\label{simulations}

\begin{figure*}[t]
\centerline{
\psfig{figure=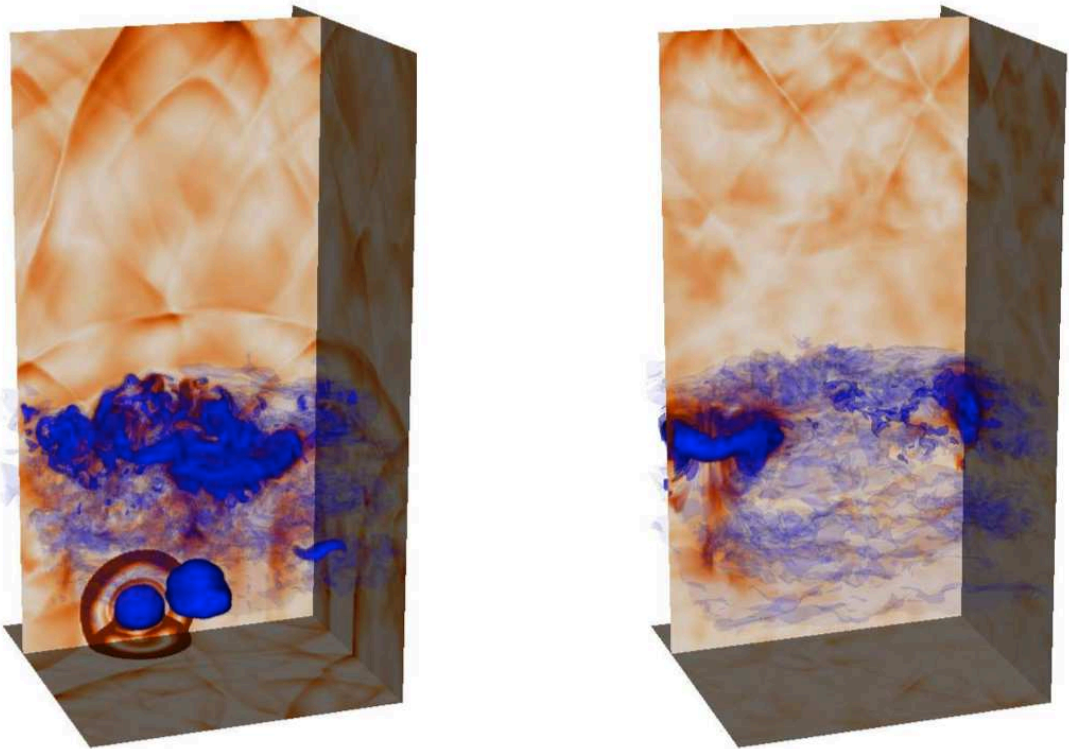,angle=90,width=0.9\textwidth}
}
\vspace{-1cm}
\caption{3D renderings of the lower half ($z<4h$) of the domain in the fiducial case ($z_0=10h$) at two times, $t=84h/c_{\rm iso,0}$ (left) and $t=180h/c_{\rm iso,0}$ (right).   We show the magnitude of the full velocity as orange shading on representative slices.   Shown in blue is a volume rendering of the injected tracer $\mu_1$, which flags the region polluted by the debris of the AGN-blown bubbles.}
\vspace{0.5cm}
\label{fig:run1_vel_tr1_3d}
\end{figure*}

Our goal is to examine the basic physical ingredients of the AGN-driven turbulent heating scenario rather than construct a realistic global model of a turbulently heated, radiatively-cooling ICM atmosphere.  Thus, we abstract the problem to one of a plane-parallel atmosphere that, at the initial time, is in hydrostatic equilibrium with a time-invariant gravitational field ${\bf g}$.    Then, working within the framework of ideal-gas hydrodynamics, we mimic AGN jet driving by detonating thermal bombs close to the base of the atmosphere at random time intervals. At some given point in time, we stop the driving and study the decay of the motions (turbulent or otherwise) within the atmosphere.  

We adopt a Cartesian coordinate system $(x,y,z)$.  Neglecting self-gravity of the ICM plasma, we assume a (fixed) gravitational field of the form
\begin{equation}\label{eqn:g}
{\bf g}=-\frac{g_0}{1+z/z_0}\hat{\bf z}.
\end{equation}
At time $t=0$, the atmosphere is isothermal  and in hydrostatic equilibrium,
\begin{equation}\label{eqn:hse}
\nabla P = -\rho {\bf g},
\end{equation}
where the pressure $P$ and the density $\rho$ are related by the (initially uniform) isothermal sound speed $c_{\rm iso,0}$ in the usual manner, $P=c_{\rm iso,0}^2\rho$.  Thus, with our chosen form for the gravitational field, the initial density profile satisfying equation~(\ref{eqn:hse}) is
\begin{equation}\label{eqn:initial_rho}
\rho=\rho_0\left(1+\frac{z}{z_0}\right)^{-z_0/h},
\end{equation}
where $h=c_{\rm iso,0}^2/g$ is the pressure scale height at the base ($z=0$) of the atmosphere.  Equation~(\ref{eqn:initial_rho}) tends to the usual exponential atmosphere as $z_0\rightarrow\infty$.  With $P=(\gamma-1)u$, where $u$ is the internal energy, we find that the \bv frequency at the initial time is
\begin{eqnarray}
N=\frac{g_0(\gamma-1)^{1/2}}{c_{\rm iso,0}\gamma^{1/2}}\left(1+\frac{z}{z_0}\right)^{-1}.
\end{eqnarray}
Hence, the \bv\ frequency decreases more rapidly, leading to stronger trapping of $g$-modes, if we consider smaller values of $z_0$.       

We use the PLUTO code \citep{mignone:07a} to model the dynamics of this atmosphere subject to AGN driving.  PLUTO evolves the equations of compressible ideal hydrodynamics in conservative form,
\begin{eqnarray}
\frac{\partial \rho}{\partial t}+\nabla\cdot(\rho\vvel)&=&0,\\
\frac{\partial}{\partial t}(\rho\vvel)+\nabla\cdot(\rho\vvel\vvel+P{\cal I})&=&-\rho\nabla\Phi,\\
\frac{\partial}{\partial t}(E+\rho\Phi)+\nabla\cdot\left[(E+P+\rho\Phi)\vvel\right]&=&0,\\
\nonumber
\end{eqnarray}
where $\vvel$ is the fluid velocity, ${\cal I}$ is the unit rank-two tensor, $E$ is the total energy density of the fluid,
\begin{equation}
E=u+{1\over 2}\rho |\vvel|^2,
\end{equation}
and $\Phi$ is the gravitational potential, which for our problem is, from equation~(\ref{eqn:hse}),
\begin{equation}
\Phi=g_0z_0\ln\left(1+\frac{z}{z_0}\right).
\end{equation}
Note that we evolve the total energy equation in conservative form --- apart from losses at any open boundaries and energy that is explicitly injected into the system, energy is conserved to machine accuracy and hence the dissipation of turbulence by grid viscosity will be captured as heating.

Driving occurs in the form of thermal bombs close to the base of the atmosphere that occur randomly in time according to Poisson statistics with a mean recurrence time $t=t_{\rm inj}$.   When such an AGN driving event is triggered, we instantaneously increase the thermal pressure in a small sphere of radius $r=r_{\rm bub}$ centered at $(x_{\rm bub}, y_{\rm bub}, z_{\rm bub})$ by an amount $\Delta P_{\rm bub}$.  We stop the driving and allow the atmospheric dynamics to decay after $t=t_{\rm stop}$.     For all simulations in this paper, we use 
\begin{eqnarray}
t_{\rm inj}&=&10h/c_{\rm iso,0},\\ 
t_{\rm stop}&=&200h/c_{\rm iso,0},\\ 
r_{\rm bub}&=&0.1h,\\
z_{\rm bub}&=&0.2h,\\
\Delta P_{\rm bub}&=&5\rho_0c_{\rm iso,0}^2. 
\end{eqnarray}
The lateral location ($x_{\rm bub}$ and $y_{\rm bub}$) of each detonation is chosen from a uniform random distribution.

Our computational domain is the Cartesian box defined by $x\in [-h,h], y\in [-h,h]$, and $z\in [0,8h]$ which we cover with $n_x\times n_y\times n_z=256\times 256\times 1024$ computational zones.  We use PLUTO in its dimensionally-unsplit mode using the {\tt hllc} Riemann solver, linear spatial interpolation, and RK2 time-stepping.   We employ periodic boundary conditions in the $x$ and $y$ directions and reflecting boundary conditions at the bottom of the atmosphere ($z=0$).  At the top of the atmosphere ($z=8h$), the boundary conditions on $P$ and $\rho$ are given by extrapolating an isothermal hydrostatic equilibrium solution into a ghost zone, whereas the velocity is set by a standard zero-gradient outflow ($v_z\ge 0$) condition.

\begin{figure*}[t]
\vspace{-2cm}
\centerline{
\psfig{figure=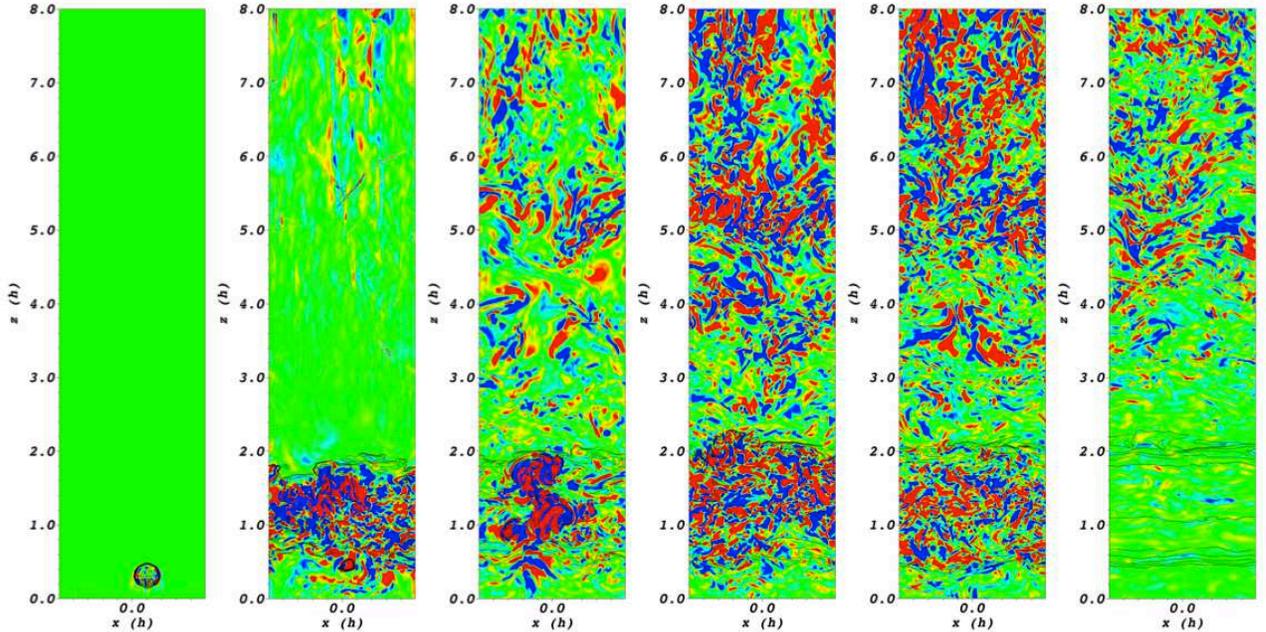,angle=90,width=\textwidth}
}
\vspace{-2.5cm}
\caption{Magnitude of $\omega_z$, the $z$-component of vorticity, in the $y=0$ plane at times $t=6, 84, 160, 180, 220,$ and $400\,h/c_{\rm iso,0}$ (from left to right).  The color table is set so that green is $\omega_z=0$, and red/blue saturate at $\omega_z=\pm 0.2$. }
\label{fig:run1_vortz_tr1_2d}
\end{figure*}

We record the full state of the atmosphere every $\Delta t=2h/c_{\rm iso,0}$.  In addition to the standard hydrodynamic variables, we also keep track of two additional passive scalar tracers, $\mu_1({\rm x},t)$ and $\mu_2({\rm x},t)$, which evolve according to
\begin{equation}
\frac{\partial}{\partial t}(\rho\mu_i)+\nabla\cdot(\rho\mu_i\vvel)=0.
\end{equation}
The first tracer $\mu_1$ is injected along with the additional heat into the thermal bombs and hence gives us a view of the region that has been directly ``polluted'' by the debris of the AGN-driven cavities --- in real systems, this would be the region containing the relativistic particles that started out in the AGN jet/cocoon if we were to ignore the diffusive transport of these particles.  The second tracer $\mu_2$ is laid down at time $t=0$ into thin horizontal layers at $z=nh/2$, $n\in \mathbb{N}$.   The evolution of this tracer allows us to gain a quasi-Lagrangian viewpoint of perturbations and disturbances within the atmosphere, at least until layers are mixed together.  

We consider three model atmospheres distinguished by the value of $z_0$.  Our fiducial case has $z_0=10h$.  In this model, $g$-modes are only weakly trapped.  We also consider cases of strong ($z_0=2h$) and very strong ($z_0=h$) trapping.  For all three models, we choose $\gamma=5/3$.   Each model was run on 1024-cores of the {\tt Deepthought2} cluster at the University of Maryland, College Park, with a run time of approximately 30,000 core hours per model.\\

\section{Results for the Weak-Trapping Case}\label{fiducial}

\subsection{Basic Dynamics and Evolution}\label{fiducial_basics}

We begin by discussing the evolution of the weak-trapping case ($z=10h$).  Following each injection event, the strong overpressure within the heated sphere drives the supersonic inflation of a low-density bubble, launching a shock that starts strong and rapidly weakens (an example is shown in the left panel of Fig.~\ref{fig:run1_vel_tr1_3d}).  The rapid expansion of the bubble stops when it reaches approximate pressure balance with its surroundings, after which point it begins to rise buoyantly.  As is well known \cite[e.g., ][]{bruggen:02a,friedman:12a}, the buoyant rise of the bubble induces Kelvin-Helmholtz and Richtmyer-Meshkov instabilities, which transform the bubble into a vortex ring followed by a turbulent wake (right panel of Fig.~\ref{fig:run1_vel_tr1_3d}).  As the bubble-ring rises, it mixes with its lower-entropy surroundings and eventually stops rising when its entropy has been reduced to the background level.  The repeated generation of these buoyant bubbles quickly forms a turbulent layer of bubble-debris.  The mixing process for the later generations of bubbles is more effective because they are rising into an already turbulent region.  In our fiducial model, the mixing and equilibration occur below $z=2h$ and the bubble-debris (as defined by tracer $\mu_1$) always remains confined to this region.  From hereon, we shall refer to this region as the bubble-debris layer.    From early times, the bubble-debris layer can be seen oscillating at approximately the \bv frequency.  

Above $z=2h$, the atmosphere remains pristine in the sense that it is unpolluted by the bubble debris.   However, it is still agitated by the weak shocks and sound waves from the AGN driving as well as by the buoyant oscillations of the bubble-debris zone below.  A central question of this study is whether this pristine region is driven into a state of turbulence.  We assess this by examining the vorticity, $\omega$.  As discussed in Section~\ref{theory}, we expect $g$-modes in the atmosphere to generate $x$- and $y$-components of vorticity, but, at linear order in the velocity perturbations, not the $z$-component $\omega_z$.  Small-scale vorticity in all directions will be produced, however, by non-linear interactions characterizing turbulence.   Thus, the production of small-scale and volume-filling $\omega_z$ is a useful probe of the development of turbulence.  The values of $\omega_z$ on a slice in the $(x,z)$-plane at representative snapshots in time are shown in Fig.~\ref{fig:run1_vortz_tr1_2d}.  At very early times (Fig.~\ref{fig:run1_vortz_tr1_2d}a), we see the generation of $z$-vorticity during the rise of an individual bubble.  Later (Fig.~\ref{fig:run1_vortz_tr1_2d}b), there is significant vorticity generation by the turbulence within the bubble-debris layer but very little in the ambient medium (apart from a few streaks of vorticity high in the atmosphere formed when weak shocks intersect).  Then, between times $t\approx140-220h/c_{\rm iso,0}$ (Fig.~\ref{fig:run1_vortz_tr1_2d}c--e), there is a volume-filling generation of small scale $\omega_z$ in the ambient medium.   This is a strong indication for the generation of volume-filling turbulence in the ambient medium.\\

\subsection{Decay of $g$-modes and Transition to Turbulence}\label{fiducial_transition}

\begin{figure}
\centerline{
\hspace{-1cm}
\psfig{figure=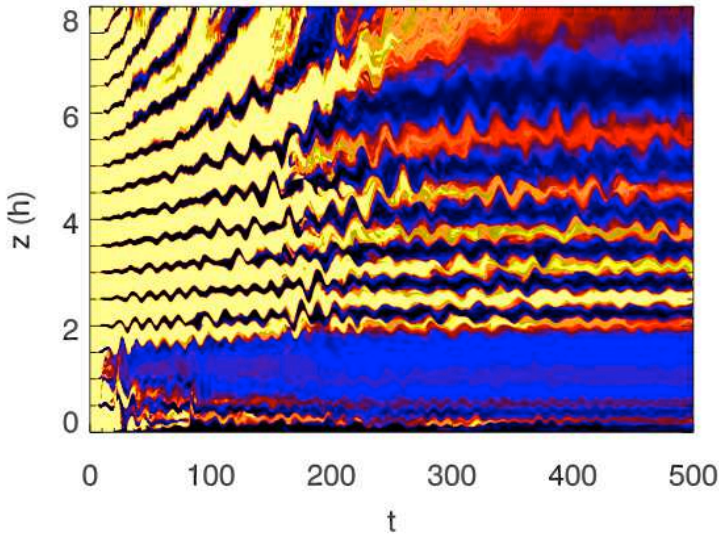,angle=90,width=0.6\textwidth}
}
\vspace{-0.5cm}
\caption{Space-time plot of a ``core sample'' ($x=y=0, z\in [0,8h]$) of the tracer $\mu_2$.  This provides a quasi-Lagrangian view of the dynamics.  We use a logarithmic color table spanning the range $\mu_2=10^{-4}$ (light yellow) to $\mu_2=1$ (black).  Time is measured in units of $h/c_{\rm iso,0}$.}
\label{fig:run1_tr2_timevar}
\end{figure}

\begin{figure}
\centerline{
%\hspace{0.3cm}
\psfig{figure=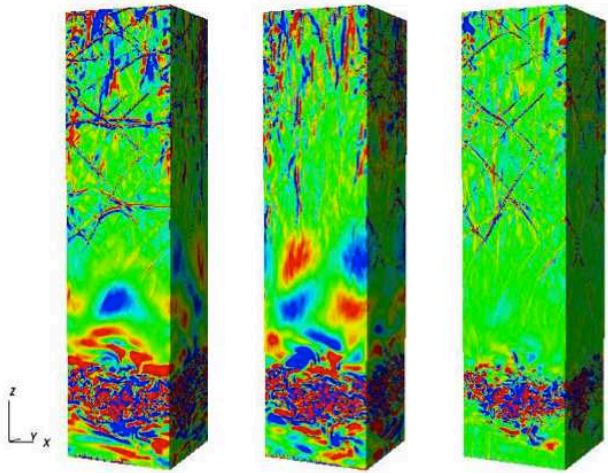,angle=90,height=0.3\textheight}
}
\caption{The three components of vorticity ($\omega_x$, $\omega_y$, $\omega_z$ from left to right) for a snapshot of the domain in the fiducial case at $t=140h/c_{\rm iso,0}$. The color table is set so that green is $\omega_{x,y,z}=0$, and red/blue saturate at $\omega_{x,y,z}=\pm 0.2$.  }
\label{fig:run1_vort}
\end{figure}

The transition to turbulence can be seen by studying the time-evolution of the passive tracer $\mu_2$. In Fig.~\ref{fig:run1_tr2_timevar}, we consider a ``core sample'' drilled though the middle ($x=y=0$) of the simulated atmosphere, and show the variation of the passive tracer $\mu_2$ as a function of height and time.   The tracer layers in the lower regions of the atmosphere ($z<2h$) are rapidly mixed by the action of the buoyant bubbling.  We also see that the highest parts of the atmosphere (those with initial positions above $z=6h$) are rapidly pushed towards the top boundary of the simulation domain due to the heating and slight resultant expansion of the  bubble-debris.  At intermediate heights, however, we clearly see the onset of oscillatory motions.  Determining the frequency of these oscillations by eye, we find good agreement (at the 10--20\% level) with the \bv frequency as given by equation~(\ref{eq:bv}).   Examining maps of the three components of the vorticity (e.g. Fig.~\ref{fig:run1_vort}), we find that these oscillations are associated with the production of vorticity in the $(x,y)$-plane but not in the $z$-direction.  Thus, we are secure in identifying them as $g$-modes.

\begin{figure*}[t]
\vspace{-2cm}
\centerline{
\psfig{figure=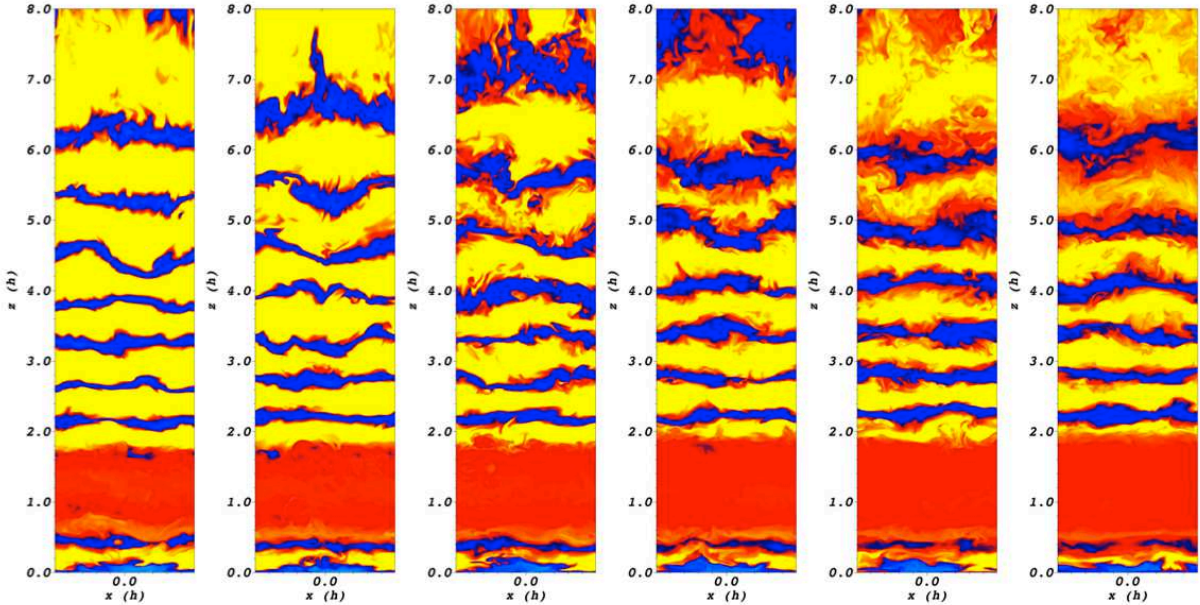,angle=90,width=\textwidth}
}
\vspace{-2.5cm}
\caption{Magnitude of the tracer $\mu_2$ on $y=0$ slices through the domain in the fiducial model at times $t=140,160,180,200,220,$ and $240h/c_{\rm iso,0}$.  A logarithmic color table is employed spanning the range $\mu_2=10^{-4}$ (yellow) to $\mu_2=1$ (dark blue). }
\vspace{1cm}
\label{fig:run1_tr2_2d}
\end{figure*}

From Fig.~\ref{fig:run1_tr2_timevar}, we clearly see that the $g$-modes undergo a transition to turbulence, at which point the $\mu_2$ layers start to mix with the surrounding ICM.   Another view of this process is presented in Fig.~\ref{fig:run1_tr2_2d}, which shows the tracer $\mu_2$ across $(x,z)$-slices in snapshots of time that span the transition to turbulence.  At the beginning of this sequence ($t=140h/c_{\rm iso,0}$; Fig.~\ref{fig:run1_tr2_2d}a), the $\mu_2$ layers are still well-defined and display a distinct sine-wave profile --- it is clear that the dominant $g$-modes have the largest non-trivial horizontal wave-numbers that will fit within our domain.  As time progresses, the $g$-modes ``break'' (see, e.g., the $\mu_2$ layer at $z=5.7h$ in Fig.~\ref{fig:run1_tr2_2d}c) and turbulence is driven.   From approximately $t=220h/c_{\rm iso,0}$ onwards, turbulence as signaled by mixing of $\mu_2$ layers and the presence of small-scale $\omega_z$ is present at all altitudes in the atmosphere.

Thus, at least in our fiducial atmosphere, we have verified two of the central tenets of the turbulent heating scenario --- that the AGN activity can indeed drive turbulence in the ambient ICM and that it is the non-linear decay of $g$-modes that gives rise to this turbulence.   Having established these basic properties of our model atmosphere, we turn to a quantitative discussion of its energetics.  

\subsection{Comparison of Compressible and Incompressible Motions: Inefficiency of AGN-Driven Turbulence}\label{fiducial_comparison}

The dynamics of our atmosphere involves both compressible (weak shocks and sound waves) and incompressible ($g$-modes and turbulence) motions.   To enable a quantitative study of the energetics, we need to decompose the velocity field into its compressible and incompressible components,
\begin{equation}
\vvel=\vvel_c+\vvel_i,
\end{equation}
where
\begin{eqnarray}
\vvel_c&=&\nabla\psi,\\
\vvel_i&=&\nabla\times {\bf a},
\end{eqnarray}
and we determine the scalar field $\psi({\rm r})$ by solving the Poisson equation,
\begin{equation}
\nabla^2\psi=\nabla\cdot \vvel.
\end{equation}
(The Cartesian nature of our simulation and the periodicity in the $x$ and $y$ directions greatly simplify the decomposition by allowing us to utilize standard Fast Fourier Transform methods.)   %An example of the decomposition is shown in Fig.~\ref{fig:run1_helmdecomp}.  

Once this decomposition is performed, we can examine the kinetic energy of the compressible and incompressible components for both the bubble-debris layer and the ambient atmosphere:
\begin{eqnarray}
K_{\rm bub,i/c}&=&\int_V{1\over 2}\rho |v_{i/c}^2| \,dV\hspace{1cm}(\mu_1\ge \epsilon),\\
K_{\rm amb,i/c}&=&\int_V{1\over 2}\rho |v_{i/c}^2|\,dV\hspace{1cm}(\mu_1< \epsilon),
\end{eqnarray}
where $\epsilon$ is taken to be a small number ($10^{-6}$) that, when compared with the tracer $\mu_1$, operationally divides volume into the bubble-debris layer and the ambient/unpolluted medium\footnote{We note that, due to the small but finite numerical diffusion in PLUTO, the whole computational domain becomes contaminated by very low levels ($10^{-15}$) of the tracer $\mu_1$.  However, the boundary of the true bubble-debris layer is well delineated and our discussion is insensitive to the precise choice of $\epsilon$ provided it is in the range $10^{-10}-10^{-4}$.}.

\begin{figure*}
\hbox{
\psfig{figure=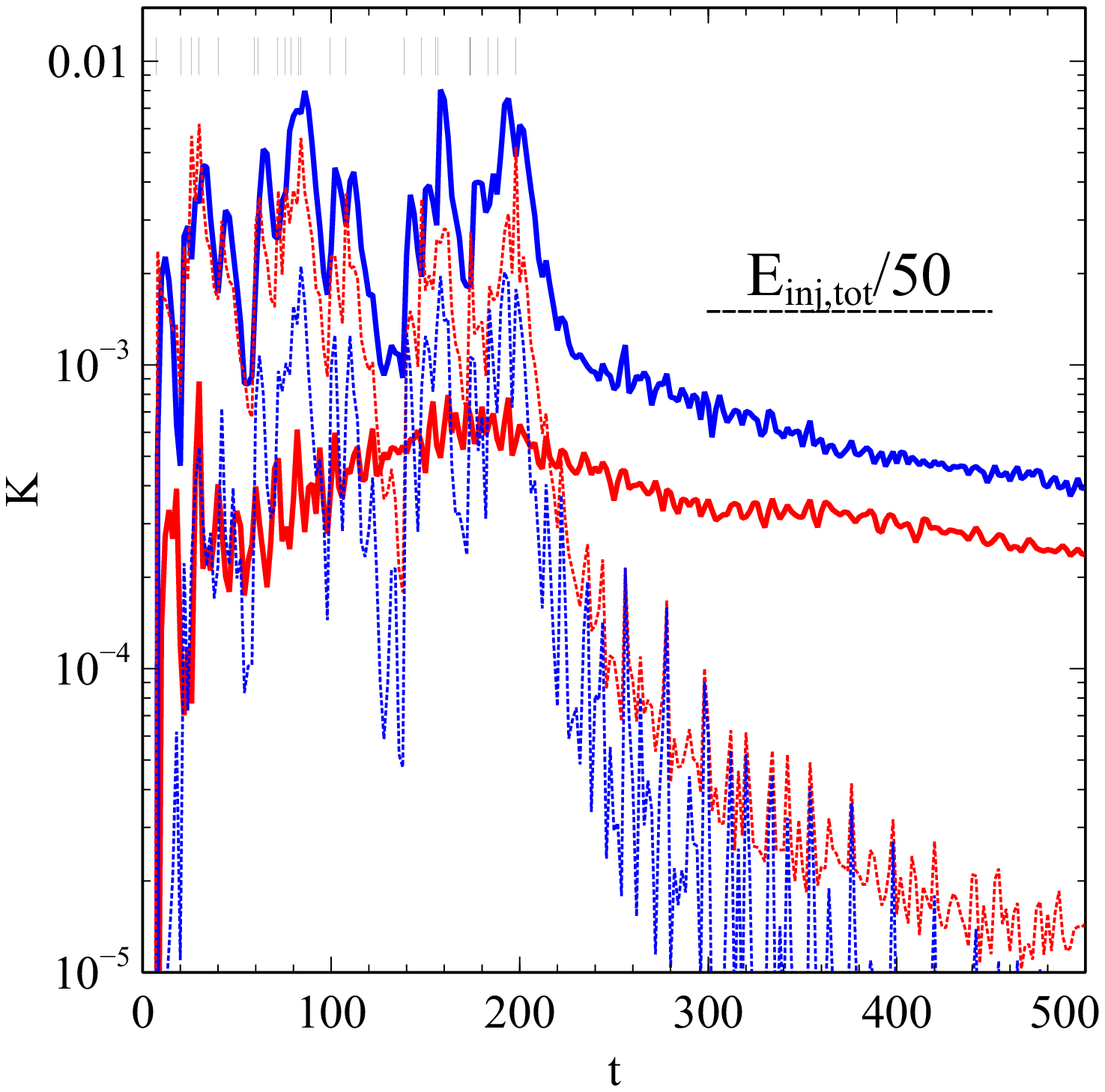,width=0.48\textwidth}
\psfig{figure=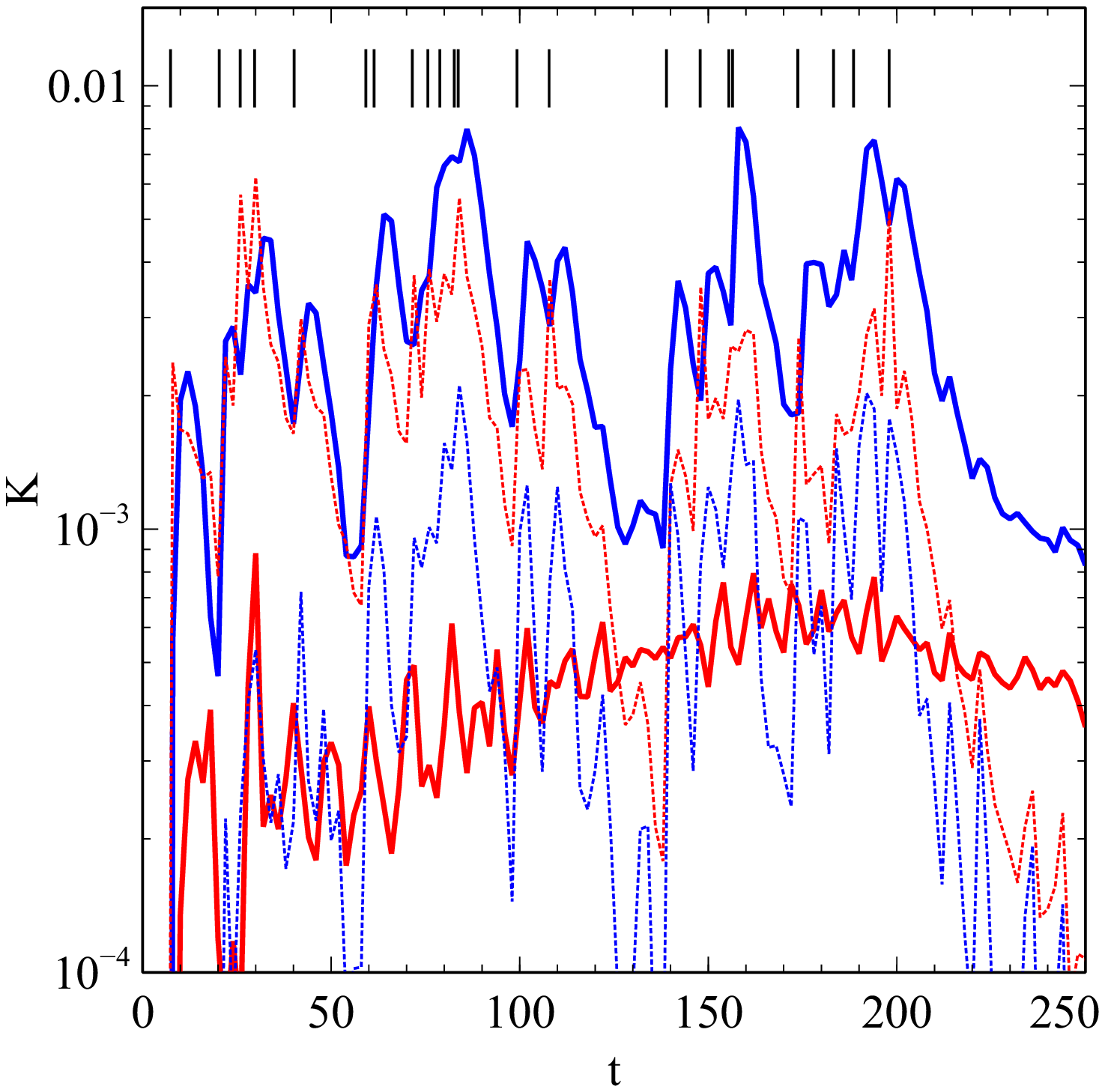,width=0.48\textwidth}
}
\caption{Decomposition of the volume-integrated kinetic energy into compressible/incompressible modes (dotted/solid lines respectively) and bubble-debris/ambient regions (blue/red lines respectively).   The right panel is a zoom in on the early phase of the evolution.  The short vertical black lines at the top of both panels show the times of the individual AGN injection events (23 events in total). Time is measured in units of $h/c_{\rm iso,0}$ and kinetic energies are in units of $\rho_{0}c_{\rm iso,0}^2h^3$.}
\label{fig:run1_ke}
\end{figure*}

Figure~\ref{fig:run1_ke} shows the compressible and incompressible components of the kinetic energy for the bubble-debris layer and ambient medium in our fiducial model.  The evolution divides into early times (during the active AGN driving phase) and late times (when the driving is shut off).   During the early/active phase, the kinetic energy of the bubble-debris layer is dominated by incompressible motions.  Combining this with the vorticity information discussed above, it is clear that this region is turbulent from an early stage.  When we compare the detailed time-evolution of the kinetic energy of incompressible motions in the bubble-debris layer with the timing of the injection events (Fig.~\ref{fig:run1_ke}b), we find that each injection deposits approximately $\Delta E=1-3\times 10^{-3}$ units of kinetic energy into the incompressible motions of the bubble ash; this amounts to between 3--10\% of the total injected energy.  The injected kinetic energy then decays on a timescale ($\Delta t\sim 10h/c_{\rm iso,0}$) that is comparable to the \bv period, suggesting that the freshly-injected turbulent motions in the bubble-debris layer are thermalized on a buoyancy timescale.    Compressible modes are always subdominant within the bubble-debris layer because they escape this relatively narrow region on a sound crossing time.

In the ambient material during this early/active phase, the situation is essentially reversed.  The kinetic energy is dominated by the compressible modes, with each injection depositing approximately $\Delta E=1-3\times 10^{-3}$ units of energy into these modes.  The incompressible modes are sub-dominant during this phase, with a kinetic energy that grows steadily with time until approximately $t=150h/c_{\rm iso,0}$, at which time it appears to saturate.  We note that this saturation time coincides with the transition to turbulence discussed in Section~\ref{fiducial_transition}.  At early times, there is direct contact between freshly injected bubbles and the ambient material, hence we see a direct correspondence between injection events and spikes in the energy of these ambient incompressible modes.  The rapid decay of these spikes likely reflects the mixing of the bubbles and ICM (and hence a reassignment of fluid elements from ambient gas to the bubble-debris layer).  At later times, a complete bubble-debris layer forms that ``shields'' the ambient ICM from the early stages of evolution of a given bubble, thus ending that correspondence.   

At late times, after the cessation of AGN driving, the kinetic energy in the compressible modes of both the bubble-debris layer and the ambient ICM quickly diminishes as the sound waves escape the upper (open) boundary (Fig.~\ref{fig:run1_ke}a).   The whole atmosphere rapidly becomes dominated by incompressible modes, with approximately twice as much kinetic energy residing in the bubble-debris.  The absolute magnitude of the kinetic energies reveal a result of central importance --- the transfer of energy from the AGN injection through to the turbulence of the ambient medium is very small, significantly less than 1\%.  

\subsection{Power Spectra of the Turbulent Ambient ICM}

\begin{figure}
\hbox{
\hspace{-2cm}
\psfig{figure=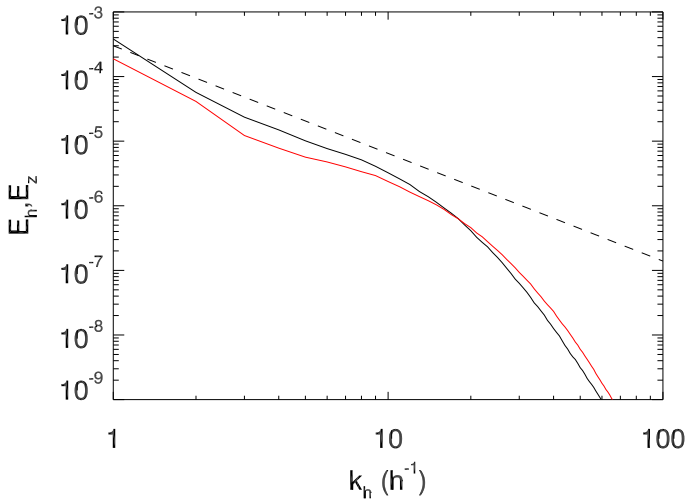,width=0.6\textwidth}
}
\caption{Power spectra of the (incompressible) horizontal motions $E_h$ (black line) and the (incompressible) vertical motions $E_z$ (red line) as a function of the horizontal wavenumber, $k_h=(k_x^2+k_y^2)^{1/2}$, for a cross section of the fiducial atmosphere $z=5h$ averaged over times $t=300-400h/c_{\rm iso,0}$.    The black dashed line shows a fiducial Kolmogorov slope of $k^{-5/3}$.  Time is measured in units of $h/c_{\rm iso,0}$ and energy power spectra are in units of $\rho_{0}c_{\rm iso,0}^2h.$ }
\label{fig:run1_powerspectra}
\end{figure}

Before leaving the fiducial case, we briefly discuss the energy power spectrum of the turbulence that we have noted.  In our quick treatment here, we only examine the dependence of these power spectra on the horizontal wavenumber $k_h=(k_x^2+k_y^2)^{1/2}$  --- we focus our attention on the power-spectrum of incompressible motions as a function of $k_h$ within horizontal cross-sections of the domain, i.e. at some given $z={\rm constant}$.   We do not discuss the dependence on the vertical wavenumber $k_z$ due to the heterogeneity and lack of periodicity in the $z$-direction    However, motivated by the expectation that the turbulence may be stratified \citep{zhuravleva:14a,zhuravleva:14b}, we examine separately the energy power spectrum of horizontal and vertical incompressible motions,
\begin{eqnarray}
E_h({\bf k})&=&\pi k_h[|\tilde{v}_{ix}({\bf k_h})|^2+|\tilde{v}_{iy}({\bf k_h})|^2],\\
E_z({\bf k})&=&\pi k_h|\tilde{v}_{iz}({\bf k_h})|^2,
\end{eqnarray}
respectively, where
\begin{equation}
\tilde{v}_{i\alpha}({\bf k_h})=\int v_{i\alpha}({\bf x})e^{i{\bf k}\cdot{\bf x}}\,d^3{\bf x},\hspace{1cm}\alpha\in \{x,y,z\},
\end{equation}
and 
\begin{equation}
{\bf k_h}=k_x{\bf \hat{x}}+k_y{\bf \hat{y}}.
\end{equation}

Figure~\ref{fig:run1_powerspectra} shows the late-time spectra spectra $E_h(k_h)$ and $E_z(k_h)$ for a cross-sectional slice of the atmosphere at $z=5h$.  This height is chosen as a representative location where the $g$-mode decay clearly drives turbulence.  Both spectra have the same shape and approximate normalization, with a steep decay on scales smaller that $0.1h$.  This scale corresponds to only $\sim 12$ computational zones and so we attribute this steepening to the effects of numerical dissipation.  On scales between $0.1-0.3h$, both power-spectra have approximately the Kolmogorov slope of $k^{-5/3}$.  The emergence of a broad power-law spectrum confirms the visual impression that a turbulent state is achieved, with very similar magnitudes of horizontal and vertical motions suggesting approximate isotropy.  For the largest scales, both power-spectra are steeper than the Kolmogorov slope which we attribute to the continuing presence of long-wavelength $g$-modes.  

\section{Behavior of the Atmosphere with $g$-mode trapping}\label{trapping}

\begin{figure*}[t]
\vspace{-3cm}
\centerline{
\psfig{figure=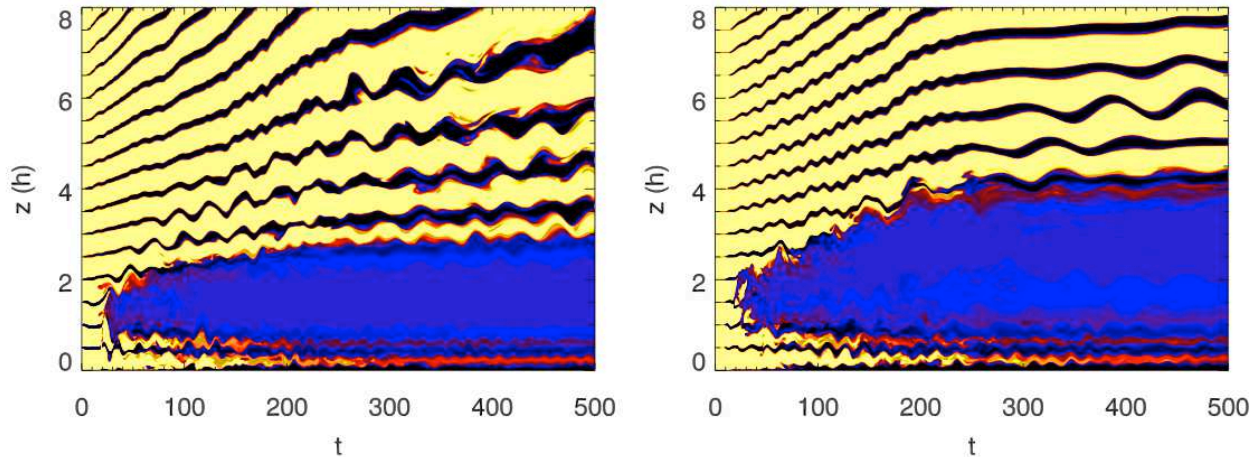,angle=90,width=\textwidth}
}
\vspace{-4cm}
\caption{Space-time plots of ``core samples'' ($x=y=0, z\in [0,8h]$) of the tracer-$\mu_2$ for the $z_0=2h$ (left) and $z_0=h$ (right) ``trapped'' cases, employing a logarithmic color table spanning the range $\mu_2=10^{-4}$ (light yellow) to $\mu_2=1$ (black).  Time is measured in units of $h/c_{\rm iso,0}$.  }
\label{fig:comparison_tr2}
\end{figure*}

We now turn to a discussion of atmospheres with strong $g$-mode trapping, i.e., our $z_0=h,2h$ cases.  One can imagine competing effects of mode trapping.  On the one hand, trapping prevents the $g$-modes from propagating out of the bubble-debris zone in the first place, diminishing the possibility of AGN-driven turbulence.  On the other hand, those modes that can propagate into the ambient medium will be confined in the lower regions of the atmosphere and hence may be locally enhanced and promote turbulence.   We will see that, at least in our model atmospheres, the first of these factors is dominant and that trapping acts to suppress the AGN driving of turbulence.  

Figure~\ref{fig:comparison_tr2} shows the time evolution of a core-sample through the $\mu_2$ tracer layers as a function of time for these two cases.  Firstly we discuss the $z_0=2h$ case.   As in the fiducial case, we see strong mixing within the bubble-debris layer which now reaches up to $z=3h$ by the end of the simulation.  The increased vertical extent of the bubble-debris layer is due to flatter entropy profile --- the mixed debris of the bubbles rise higher before equalizing to background entropy level.  The $\mu_2$ tracer layers reveal oscillations in the ambient medium starting in the lower layers and, at later times, extending to the higher layers.  As in the fiducial case, these oscillations are found to have approximately the \bv frequency and are associated with the production of vorticity in the $(x,y)$-plane, hence we associate them with $g$-modes.   In contrast to the fiducial case, there is only a very modest level of breaking of these modes.   Only very weak turbulence is generated, with most of the energy remaining in the $g$-modes by the end of the simulation.  

Many of these trends hold for the atmosphere with the strongest trapping, $z_0=h$.  The bubble-debris layer now extends to $z=4h$ due to an even flatter entropy profile, and at late times (once driving stops) the ambient upper atmosphere supports low-frequency $g$-modes that remain essentially linear through the extent of the simulation.   No turbulence is generated within the ambient atmosphere.   

\begin{figure}
\centerline{
\psfig{figure=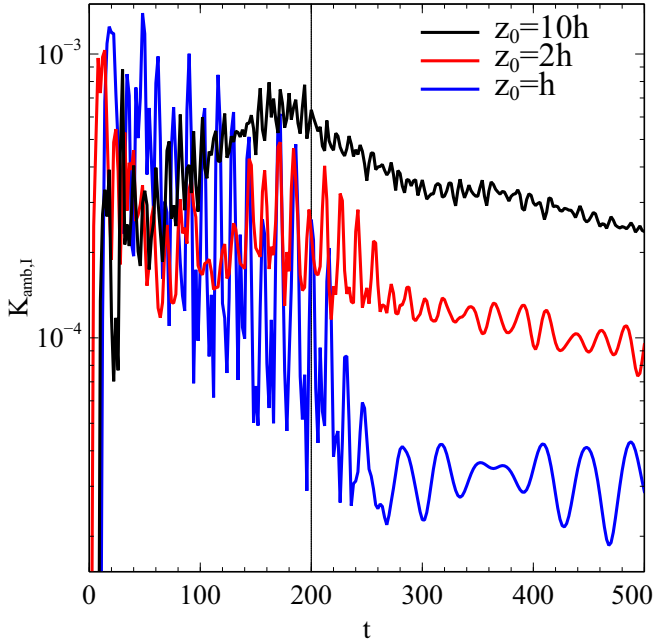,width=0.48\textwidth}
}
\caption{Volume-integrated kinetic energies associated with incompressible motions of the ambient gas for the three models presented in this paper.  Time is measured in units of $h/c_{\rm iso,0}$ and kinetic energies are in units of $\rho_{0}c_{\rm iso,0}^2h^3$. }
\label{fig:comparison_ke}
\end{figure}

The role of $g$-mode trapping is revealed most directly through an analysis of the incompressible component of the kinetic energy in the ambient matter, as shown in Fig.~\ref{fig:comparison_ke}.  During the early/driven phase, these three runs show very similar levels of incompressible dynamics.  However, once driving ceases, it is apparent that the atmosphere becomes progressively more quiescent as the trapping increases (i.e., $z_0$ decreases).  

Finally, by way of a technical comment, we note that, during the driving phase ($t<200h/c_{\rm iso,0}$) the $z_0=h$ atmosphere is seen to expand and, at all heights, to undergo high-frequency oscillations (Fig.~\ref{fig:comparison_tr2}, right panel).   These oscillations are much faster than the \bv frequency (with a period of approximately $\Delta t=12h/c_{\rm iso,0}$ compared with $2\pi/N\approx 10[1+z/h]\,h/c_{\rm iso,0}\approx 60\,h/c_{\rm iso,0}$, evaluating at $z=5h$), have no associated vorticity generation, and are coherent across heights.  We identify this as the fundamental tone of a global acoustic mode resonating within the ``cavity'' defined by our reflecting boundary condition at $z=0$ and velocity outflow (zero-gradient) boundary condition at $z=8h$.  This, of course, flags these oscillations as being artificial and an artifact of our finite computational domain.  They do, however, appear to dissipate quickly once the driving stops, leaving the $g$-modes which are of primary interest to us.   

The fact that these global acoustic oscillations are not seen in the other cases ($z_0=2h,10h$) can be understood in terms of the density profile characterizing each run.  The (initial) density ratio between the bottom and top boundaries is
\begin{equation}
\frac{\rho_{\rm top}}{\rho_{\rm bottom}}=\left(1+\frac{8h}{z_0}\right)^{-z_0/h},
\end{equation}
and, while the density profile evolves somewhat during the driving phase, it retains a strong memory of this initial state.  Evaluating the ratioes for the three cases $z_0/h=1,2,10$ gives $\rho_{\rm top}/\rho_{\rm bottom}\approx 0.11, 4.0\times 10^{-2}, 2.8\times 10^{-3}$ respectively.  For sufficiently extreme density ratios, the atmosphere at the top of the box is so rarified that the dynamics of the lower atmosphere is insensitive to the existence of the upper computational boundary (formally, the eigenmode characterizing global acoustic modes becomes insensitive to the upper boundary condition).  Our results demonstrate that the $z_0=2h$ and $10h$ cases are in this regime.  For the $z_0=h$ case, on the other hand, the upper atmosphere maintains sufficient density to enable an acoustic resonance to be established.

\section{Discussion}\label{discussion}

The goal of this work is to examine, or at least start the examination of, the basic fluid dynamics underlying the AGN driving of ICM turbulence.  Working within the framework of ideal hydrodynamics, we find that AGN activity (which we model as explosive events at the base of the atmosphere) can indeed drive $g$-modes into the atmosphere, which subsequently decay into volume-filling turbulence.   Within our model, however, this process is inefficient --- less than 1\% of the injected energy eventually ends up as the kinetic energy of turbulence in the ambient ICM, whereas the vast majority of the energy is lost in compressible modes or rapidly thermalized within the bubble-debris layer.  Strong trapping of the $g$-modes further decreases this efficiency and can even render them so weak as to not generate turbulence at the Reynolds numbers afforded by the resolution of our simulations.

Taken at face value, this inefficiency of turbulent dissipation is inconsistent with observations.  The energetics of AGN in cool core clusters are usually estimated through a determination of the product $PV$, where $P$ is the (measured) pressure of the ICM surrounding a cavity of (measured) volume $V$.  On purely energetic grounds, the AGN needs to transfer $1-20PV$ of heat per cavity to the cluster core in order to balance radiative cooling \citep{birzan:04a,panagoulia:14a}.  In our model, we find that the initial expansion phase of the bubbles (which take the bubbles from an overpressure by a factor of 6 down to pressure equilibrium with the surrounding medium) only increases their volume by a factor of 2.5--3, with approximately 70\% of the energy being used to do work on and directly shock-heat the surrounding shell of gas. Thus, examining just the energy content of a cavity will under-estimate the injected power by a factor of $\sim 3$.  However, even adopting a generous value of 1\% for the efficiency with which the AGN transfers its energy to turbulence, we still infer that the AGN transfers only $\sim 0.03PV$ per cavity to the ICM turbulence.
 
Of course, our model is highly idealized in order to facilitate a detailed study of the hydrodynamics, and we must critically assess the degree to which the idealizations affect our results.   

Firstly, real clusters are spherical, not plane-parallel, and geometry is always important for the detailed dynamics.  For example, in a spherical cluster with a realistic radial profile of pressure, acoustic modes generated at the center will weaken (i.e., $\delta p/p$ will decrease) as the wave propagates outwards and carries an energy flux that diminishes as $1/r^2$.  In our plane-parallel atmospheres with periodic boundary conditions in the $x$- and $y$-directions, on the other hand, the upper atmosphere is seeing quasi-planar acoustic modes which can re-steepen and become shocks.  Indeed, examining the entropy of our $\mu_2$ tracer layers in the fiducial case reveals heating of the uppermost regions of the atmosphere that we attribute precisely to the action of steepened acoustic modes.

Secondly, we have adopted the framework of inviscid ideal hydrodynamics.  One consequence of this is that we may underestimate how long the structural integrity of the buoyantly rising bubbles is preserved \citep{reynolds:05a}, which may, in turn, diminish the transference of energy from the bubble-induced upwell into the $g$-modes.  Magnetic fields can play an important role here.  While the magnetic field in the bulk of the ICM is expected to be weak (with magnetic pressures that are approximately 1\% of the gas pressure), the draping of field lines across the rising-cap of a buoyant bubble may significantly reduce the disruption of the bubble and subsequent mixing with the ambient ICM \citep{ruszkowski:07a,dursi:08a}, potentially impacting the efficiency with which the AGN drives ICM turbulence.  Even if magnetic forces are weak enough so as to be unimportant, the magnetic field strongly affects transport processes, introducing a very strong anisotropy into the viscous stress tensor and thermal conductivity.  In general terms, the anisotropic viscosity and conduction can affect the convective stability of the ICM \citep{balbus:00a,quataert:08a,balbus:10a,kunz:11a,kunz:12a} and both the conduction and the viscosity can dramatically enhance the dissipation of compressible modes.  Of more direct interest to the results of this paper, the combined action of the anisotropic conduction and the radiative cooling can drive $g$-modes to be overstable \citep{balbus:10a}; this maybe be one possible route for increasing the efficiency of energy transfer from the AGN to the ambient ICM.

\section{Conclusions}\label{conclusions}

We have constructed model atmospheres tailored to test the physical assumptions underlying the AGN-induced turbulent heating model of cooling-core clusters.  We find that AGN-like disturbances can indeed induce volume-filling turbulence in the ambient ICM via the decay of $g$-modes.  In our ideal hydrodynamic model, however, we find that the transfer of energy from the AGN-injection process into the turbulence of the ambient gas is quite inefficient (less than 1\% for our fiducial models).  For the models assessed here, the trapping of $g$-modes acts to decrease the efficiency further because the excitation of turbulence in our model relies on $g$-mode propagation from the cluster center where they are launched by explosive events to the bulk of the ICM core.  We stress, however, that this is just the first step in the theoretical assessment of the AGN-driven turbulent heating model and certainly not the last word.  Next steps must examine models that include the action of radiative cooling, magnetic fields, (anisotropic) thermal conduction and (Braginskii) viscosity.  While these are ingredients that, from first principles, {\it should be present}, we advocate a systematic and controlled study of models of increasing complexity.  Only then can the essential underlying physical mechanisms be revealed.   

\medskip

CSR is grateful for financial support from the Simons Foundation (through a Simons Fellowship in Theoretical Physics), a Sackler Fellowship (hosted by the Institute of Astronomy, Cambridge), and the National Science Foundation (under grant AST1333514).  SAB acknowledges support from the Hintze Family Charitable Foundation, and from the Royal Society in the form of a Wolfson Research Merit Award.  CSR thanks the Department of Physics at the University of Oxford and the Institute of Astronomy at the University of Cambridge for their hospitality during extended visits while this work was being performed.  The simulations presented in this paper were performed on the {\tt Deepthought2} cluster, maintained and supported by the Division of Information Technology at the University of Maryland College Park.

\bibliographystyle{jwapjbib}
%\bibliography{chris}

\end{document}